\documentclass[journal]{IEEEtran}
\usepackage{amsmath,amsfonts,graphicx,tabularx,color}
\usepackage{amssymb,textcomp,mathtools,amsthm,gensymb}
\usepackage{bm,upgreek,algorithm,hyperref}
\usepackage{multirow,booktabs,hhline,array}
\usepackage{cite,url,makecell,setspace,listings}
\RequirePackage{doi}

\definecolor{codegreen}{rgb}{0,0.6,0}
\definecolor{codegray}{rgb}{0.5,0.5,0.5}
\definecolor{codepurple}{rgb}{0.58,0,0.82}
\definecolor{backcolour}{rgb}{0.95,0.95,0.92}

\lstdefinestyle{mystyle}{
    backgroundcolor=\color{backcolour},
    commentstyle=\color{codegreen},
    keywordstyle=\color{magenta},
    numberstyle=\tiny\color{codegray},
    stringstyle=\color{codepurple},
    basicstyle=\ttfamily\scriptsize,
    breakatwhitespace=false,
    breaklines=true,
    captionpos=b,
    keepspaces=true,
    numbers=left,
    numbersep=5pt,
    showspaces=false,
    showstringspaces=false,
    showtabs=false,
    tabsize=2
}

\theoremstyle{definition}

\theoremstyle{plain}

\theoremstyle{remark}

\title{Fast Random Approximation of Multi-channel Room Impulse Response}

\author{Yi~Luo*, Rongzhi Gu*\thanks{*Equal contribution.}}

\begin{document}
\maketitle
\setlength{\abovedisplayskip}{2pt}
\setlength{\belowdisplayskip}{2pt}
\setlength{\abovedisplayshortskip}{2pt}
\setlength{\belowdisplayshortskip}{2pt}

\begin{abstract}
Modern neural-network-based speech processing systems are typically required to be robust against reverberation, and the training of such systems thus needs a large amount of reverberant data. During the training of the systems, on-the-fly simulation pipeline is nowadays preferred as it allows the model to train on infinite number of data samples without pre-generating and saving them on harddisk. An RIR simulation method thus needs to not only generate more realistic artificial room impulse response (RIR) filters, but also generate them in a fast way to accelerate the training process. Existing RIR simulation tools have proven effective in a wide range of speech processing tasks and neural network architectures, but their usage in on-the-fly simulation pipeline still remains questionable due to their computational complexity or the quality of the generated RIR filters. In this paper, we propose FRAM-RIR, a fast random approximation method of the widely-used image-source method (ISM), to efficiently generate realistic multi-channel RIR filters. FRAM-RIR bypasses the explicit calculation of sound propagation paths in ISM-based algorithms by randomly sampling the location and number of reflections of each virtual sound source based on several heuristic assumptions, while still maintains accurate direction-of-arrival (DOA) information of all sound sources. Visualization of oracle beampatterns and directional features shows that FRAM-RIR can generate more realistic RIR filters than existing widely-used ISM-based tools, and experiment results on multi-channel noisy speech separation and dereverberation tasks with a wide range of neural network architectures show that models trained with FRAM-RIR can also achieve on par or better performance on real RIRs compared to other RIR simulation tools with a significantly accelerated training procedure. A Python implementation of FRAM-RIR is available online\footnote{\href{https://github.com/tencent-ailab/FRA-RIR/tree/fram_rir}{https://github.com/tencent-ailab/FRA-RIR/tree/fram\_rir}}. 
\end{abstract}


\section{Introduction}
\label{sec:intro}
Reverberation occurs in everyday conversations when speakers or other sound sources are in an indoor environment. To ensure the robustness of recent neural speech processing and modeling systems under such scenarios, artificial reverberation is typically utilized in the training of the systems to enlarge the number of available training data and the coverage of the data across different room conditions \cite{kim2017generation}. The method to simulate artificial reverberation, typically depicted by a room impulse response (RIR) filter, is thus required to mimic the pattern of realistic reverberation. Moreover, with more and more modern systems select \textit{on-the-fly} data simulation pipeline, which allows the generation of infinite number of training data without pre-generating and saving then on harddisk and is able to improve the system performance in a wide range of tasks \cite{nguyen2020improving, zhang2021ft, zeghidour2021wavesplit, subakan2021attention, lam2021fly}, the simulation speed of the RIR filters also becomes important in the training pipeline.

Multiple RIR simulation tools have been investigated and proposed in the past years to improve the simulation quality and accelerate the simulation process \cite{valimaki2012fifty, valimaki2016more}. Physical-modeling based methods have been the mainstream, as they can accurately model the sound reflections given the room conditions. One of the most popular methods is the  \textit{image-source method (ISM)} \cite{allen1979image}, where the sound paths are calculated by mirroring the original sound sources by the room boundaries (e.g., floors and walls). To accelerate the simulation process and improve the quality of the generated filter, diffuse-based methods have been proposed to bypass the explicit calculation of late reverberation sound paths \cite{lehmann2009diffuse, tang2020improving}. ISM-based algorithms and tools have been widely used in the training of most of the neural network speech processing systems and proven effective in a wide range of tasks such as automatic speech recognition (ASR) \cite{couvreur2000use, tashev2005reverberation, chang2019mimo}, speech enhancement \cite{zhao2017two, zhao2020noisy}, speech separation \cite{maciejewski2020whamr, aralikatti2021improving}, and speech dereverberation \cite{han2015learning, wu2016reverberation}. Beyond ISM-based tools, ray-tracing-based methods have also been explored to support the modeling of more realistic rooms to generate better RIR filters \cite{funkhouser2004beam, scheibler2018pyroomacoustics}. Stochastic approximations such as velvet noise and rescaled impulse train have also been used to simulate artificial reverberation in a faster way \cite{jarvelainen2007reverberation, valimaki2013perceptual, valimaki2016more, masztalski2020storir}. With the recent development of generative models, neural networks have also been utilized to refine simulated RIR filters to better match the distributions of the real-recorded RIR filters \cite{ratnarajah2020ir, ratnarajah2021ts, ratnarajah2022fast}. 

Although the aforementioned methods have been successfully applied in various tasks and applications, their usage in the training of modern neural-network-based speech processing systems still has several difficulties. ISM-based methods typically assumes an empty rectangular or parallelepiped room, and such assumption may cause the ``sweeping echo effect'' which hurts the models' generalization ability in real-world when trained with such RIRs \cite{kiyohara2002sweeping, de2015modeling}. To alleviate the sweeping echo effect, randomizing the virtual sound source locations has shown effective under various statistical assumptions \cite{hanyu2012new, de2015modeling}, while the explicit calculation of the sound paths is still required. The calculation of the sound paths can also be time-consuming and need further acceleration with certain hardware such as GPUs \cite{fu2016gpu, diaz2020gpurir}, which makes them harder to run in parallel when fast data simulation is required in distributed model training. Ray-tracing-based methods can be computationally heavy, and to simulate realistic RIR filters one may need to perform room modeling in advance, which is still not suitable for on-the-fly data generation. Stochastic approximations or noise reshaping methods can generate RIR filters in a fast way, but the filters typically behave different from real RIR filters, and the methods themselves are in general hard to be generalized to multi-channel scenario where the time delay of arrival (TDOA) of each source-microphone pair need to be accurately preserved. Neural-network-based methods can be fast when the size and the complexity of the neural networks are acceptable, however it is relatively harder to perform fine-grained control on the generated RIR filters, e.g., to ensure that the generated RIR filters accurately preserves the desired T60 or direction-of-arrival (DOA) of each sound source. These drawbacks make existing methods hard to be applied in the on-the-fly training pipeline for neural network models.

In this paper, we extend our previous work on fast simulation of single-channel RIR filters \cite{luo2022fra} and propose \textbf{\textit{F}}ast \textbf{\textit{R}}andom \textbf{\textit{A}}pproximation of \textbf{\textit{M}}ulti-channel \textbf{\textit{RIR}} (\textbf{\textit{FRAM-RIR}}), a simple method to simulate multi-channel RIR filters with accurate DOA information for the sound sources which is fast enough to perform on-the-fly data simulation. FRAM-RIR follows the problem definition of standard ISM-based algorithms, but approximates the explicit calculation of sound paths by randomly sampling the location and number of reflections of each virtual sound source based on several heuristic assumptions. Such sampling process can be done in parallel for all virtual sound sources, which enables FRAM-RIR to be fast and suitable for parallel processing on CPUs. With a standard desktop-level CPU, FRAM-RIR can generate realistic RIR filters within 140~ms on a single CPU thread when generating RIR filters for a 4-mic array, up to 40 times faster than existing ISM-based tools. Moreover, visualizations on oracle beampatterns and directional features of speech convolved with real RIR filters and different simulated RIR filters show that FRAM-RIR can better mimic realistic spatial patterns compared to other tools. Experiment results on multi-channel noisy speech separation and dereverberation tasks with a wide range of neural network architectures show that models trained with FRAM-RIR can also achieve on par or better performance on real RIRs compared to other RIR simulation tools.

The rest of the paper is organized as follows. Section~\ref{sec:ISM} provides a brief review to the ISM-based RIR simulation method. Section~\ref{sec:FRAM} introduces the proposed FRAM-RIR method. Section~\ref{sec:config} describes the experiment configurations. Section~\ref{sec:result} analyzes the behavior of FRAM-RIR and presents the results on separation and dereverberation tasks with various neural network architectures. Section~\ref{sec:conclusion} concludes the paper.

\section{Image-source Method Recap}
\label{sec:ISM}
\begin{figure*}[t]
    \centerline{\includegraphics[width=17cm]{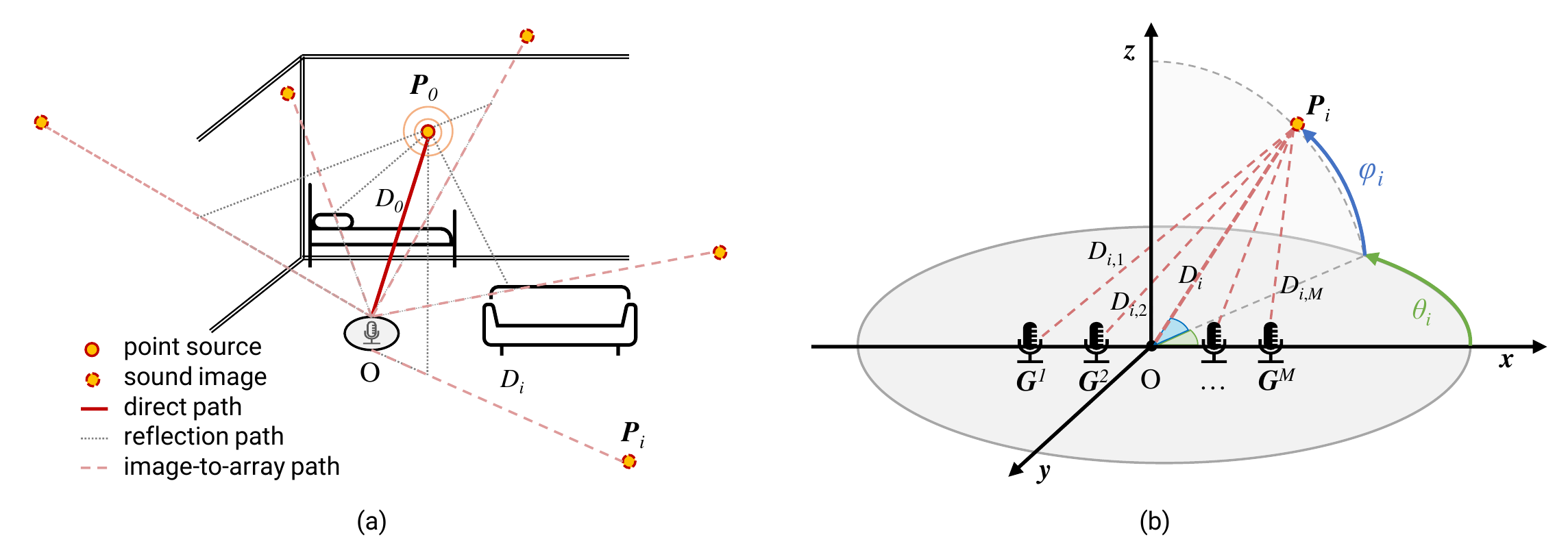}}
    \caption{(a) A point source $P_0$ and its direct path $D_0$ to the receiver (e.g., a microphone array), several early sound images $P_i$ and their corresponding propagation paths in a shoebox-shape room; (b) The $i$-th image can be uniquely identified with the image-to-receiver distance $D_i$, azimuth $\theta_i$ and elevation $\phi_i$.}
    \label{fig:images}
\end{figure*}

We adopt the definition of an image-method-generated RIR filter in \cite{kim2017generation}:
\begin{align}
\begin{split}
    h^m[n] &= \frac{1}{D_{0,m}} \delta \left[n - \left\lceil \frac{D_{0,m}f_s}{c_0} \right\rceil \right] \\
    &+ \sum_{i=1}^I \frac{r^{g_{i,m}}}{D_{i,m}} \delta \left[n - \left\lceil \frac{D_{i,m}f_s}{c_0} \right\rceil \right]
\label{eqn:image}
\end{split}
\end{align}
where $m \in [1, \ldots, M]$ denotes the receiver index, $I$ denotes the total number of virtual sound sources, $D_{0,m}$ denotes the distance of the direct-path sound source to the $m$-th receiver, $D_{i,m}$ denotes the distance from the $i$-th virtual sound image to the $m$-th receiver, $r$ denotes the reflection coefficient of the surface, $g_i$ denotes the number of the reflections of the $i$-th sound source to the $m$-th receiver, $f_s$ denotes the target sample rate, and $c_0$ denotes the sound velocity. Figure~\ref{fig:images} (a) shows the illustration of the sound images of a point source $P_0$ inside a room. We follow the same estimation of the reflection coefficient via the Eyring's empirical equation \cite{beranek2006analysis, kim2017generation}:
\begin{align}
    r = \sqrt{1 - \left(1 - e^{-0.16R/ T_{60}}\right)^2}
\label{eqn:reflect}
\end{align}
where $R$ denotes the ratio between the volume and the total surface area of the room, and $T_{60}$ denotes the reverberation time that takes for the sound to decay by 60 dB in the room.

As $h^m[n]$ is an impulse train and its sample rate affects its temporal resolution, the sample rate needs to be large enough to ensure that the TDOA of different virtual sound sources can be effectively modeled even with small-spacing microphone arrays. We thus adopt the same strategy as \cite{kim2017generation} such that $h^m[n]$ is first generated at a very high sample rate $r_hf_s$ and then downsampled to an intermediate sample rate $r_lf_s$ with $1<r_l<r_h$ being two rescaling factors, and then a high-pass filter with a cut-off frequency of 80 Hz is applied to remove the unwanted low-frequency components \cite{allen1979image, kim2017generation}. The filtered RIR filter is then downsampled again to the target sample rate $f_s$ to serve as the final output to be convolved with the actual sound source (whose sample rate is also $f_s$). In practice, we set $r_h=\lfloor 10^6/f_s \rfloor$ and $r_l=\lfloor \sqrt{r_h} \rfloor$ to balance the simulation speed and the RIR quality.

\section{Fast Random Approximation of Multi-channel RIR}
\label{sec:FRAM}
In our previous work on fast random approximation of single-channel RIR (FRA-RIR) \cite{luo2022fra}, we proposed three core modifications to the calculation of equation~\ref{eqn:image}:
\begin{enumerate}
    \item The room-related statistics $R$ was randomly sampled instead of explicitly calculated via the assumption of empty rectangular or parallelpiped rooms.
    \item The explicit calculation of $D_{i,m}$ was replaced by sampling it from a probability distribution.
    \item The explicit calculation of $g_{i,m}$ was replaced by defining it as a function of $D_{i,m}$ with random perturbations.
\end{enumerate}

While FRA-RIR achieved on par or better performance than other RIR simulation tools with a significantly faster simulation speed, a core problem for it was that its application in multi-channel scenarios was limited, as the images for different microphones need to be aligned to properly represent their spatial locations. Spatial features such as interaural phase differences (IPDs) and angle feature (AF) should also be aligned according to the correct microphone spacing. However, as all the simulations were done randomly, such properties cannot be easily achieved via the original FRA-RIR method. Here we extend FRA-RIR to support multi-channel RIR simulation while satisfying the aforementioned requirements, which we refer to as the \textit{fast random approximation of multi-channel RIR (FRAM-RIR)} method.

\subsection{Simulating Room-related Statistics}
\label{sec:room}

Different from FRA-RIR where the simulation of room-related statistics was by randomly sampling $R$ and $T_{60}$ in equation~\ref{eqn:reflect}, we step back to the original definition of $R$ which is calculated by randomly sampling a room size with the assumption of an empty rectangular or parallelpiped room. The reason we select the standard configuration of $R$ is not only because empirically it leads to on par performance as a randomly-sampled $R$, but also because the properties of a realistic room, e.g., different sound absorption coefficients of different furniture or ornaments and irregular room sizes, can be simulated by applying \emph{fractional number of reflections}, i.e., setting $g_{i,m}$ to be rational numbers instead of integers, and random perturbations to the number of reflections, which we will discuss in Section~\ref{sec:reflect}.

\subsection{Simulating Source-receiver Distances of the Images}
\label{sec:dist}

A key problem in multi-channel RIR simulation is that the TDOA between the microphones should be accurately simulated for all virtual sources. FRAM-RIR simulates the location of the virtual sound images by \textit{randomly generating their 3D coordinates inside the room}, which is done by simulating the image-to-receiver distance $D$, azimuth $\theta$, and elevation $\phi$ on a sphere whose center is a selected microphone or a predefined 3D coordinate inside the room (we drop the subscript $m$ where there is no ambiguity). Figure~\ref{fig:images} (b) shows an example of a linear microphone array with the center of the array used as the center of the sphere, where the x axis is defined by the location of the microphones. The coordinate of the $i$-th virtual sound source is sampled at the surface of the sphere with radius $D_i$, azimuth $\theta_i$, and elevation $\phi_i$, where $\theta_i$ is uniformly sampled between $[0, 2\pi]$, $\phi_i$ is uniformly sampled between $[-\pi/2, \pi/2]$, and $D_i$ is sampled by sampling the distance ratio (DR) between $D_i$ and the direct-path distance ratio $DR_i \triangleq D_i/D_0, DR_i > 1$ to ensure that all virtual sound sources have longer propagation distances than the direct sound source. We further assume that $DR_i$ can be sampled from a predefined probability distribution function (PDF). While it is difficult to accurately estimate a general PDF for virtual sound distances in various rooms or environments, here we make a simple assumption that the number of virtual sources increases as the sampled source-receiver distance increase due to the increasingly complicated reflection conditions. Hence we adopt a quadratic function as the PDF to sample the distances $D_i$:
\begin{align}
    P(x) = \begin{dcases*}
            \frac{3x^2}{\beta^3 - \alpha^3}, & $\alpha < x \leq \beta$ \\
            0, & otherwise
            \end{dcases*}
\end{align}
where $0\leq\alpha<\beta\leq1$ are scalars controlling the range of the distribution. To ensure that $DR_i > 1$, we first sample $\hat{DR}_i \in [\alpha, \beta]$ from $P(x)$ and calculate $DR_i$ by linearly rescaling $\hat{DR}_i$ to range $[1, c_0T_{60}/d_0]$:
\begin{align}
    DR_i = 1 + \frac{\alpha}{\beta-\alpha}(\frac{\hat{DR}_i}{\alpha} - 1)(\frac{c_0T_{60}}{d_0}-1)
\end{align}
where $c_0T_{60}$ is the maximum propagation distance for a virtual source. The actual propagation distance $D_i$ can then be calculated as $D_i = D_0\times DR_i$.

\subsection{Simulation Number of Reflections of the Images}
\label{sec:reflect}

Next we need to generate the number of reflections for each virtual source. We first calculate the number of reflections $RR_{max}$ by which the farthest virtual sound decays by 60 dB given the reverberation time $T_{60}$, direct-path distance $D_0$, reverberation coefficient $r$ and sound velocity $c_0$:
\begin{align}
    RR_{max} = (\text{log}_{10}\,c_0T_{60} - \text{log}_{10}\,d_0 - 3) /\text{log}_{10}\,r
\end{align}
where $c_0T_{60}$ corresponds to the longest possible propagation distance. $RR_{max}$ can thus be treated as the maximum number of reflections a virtual source may encounter, as simulated RIRs typically only consider the range within $T_{60}$. We then sample the number of reflections $g_i \in [1, RR_{max}]$ for the $i$-th virtual source by defining it as a function of $D_i$, and further add a random perturbation to it:
\begin{align}
\begin{split}
    p_i &\sim \mathcal{U}(a, b) \\
    g_i &= 1 + (\frac{D_i}{c_0T_{60}})^2 \cdot (RR_{max} - 1) + p_i \cdot DR_i^{\tau} \\
    g_i &= \text{max}(\text{min}(g_i, RR_{max}), 1)
\end{split}
\end{align}
where $\mathcal{U}$ denotes the uniform distribution, $p_i$ denotes the random perturbation on the number of reflections, and $\tau>0$ denotes the distance shrinkage factor. Here we assume that virtual sources with longer propagation distances may encounter more reflections. The perturbation term can be explained by the assumption that virtual sources with a similar overall propagation distances may also have different numbers of reflections. Moreover, note that since we allow $g_i$ to be a non-integer, such \textit{fractional number of reflections} mentioned in Section~\ref{sec:room} can be interpreted as a different reflection coefficient $\hat{r}$, defined by $\hat{r} \triangleq r\cdot e^{\frac{q_i}{\lfloor q_i \rfloor}}$, for the current virtual source with number of reflections $\lfloor g_i \rfloor$. This allows us to model more diverse and complicated source reflection patterns and mitigate the unrealistic assumption in standard ISM methods that the reflection coefficient $r$ keeps identical in all virtual sources. Here we set the perturbation to be a function of the propagation distance $D_i$ based on the heuristic assumption that virtual sources that travels longer distances may also meet surfaces with different materials, thus result in a larger variability in the number of reflections (and reflection coefficient).

\subsection{Generation of the RIR Filter}
\label{sec:RIR}

After the sampling of $D_i$ and $g_i$, $h[n]$ can be generated by summing up the rescaled impulse trains of all virtual sources. We first initialize $h[n]$ to an all-zero vector of length $L\triangleq\lceil T_{60}r_hf_s\rceil$ and then add each virtual source to $h[n]$:
\begin{align}
    q_i &= \text{min}(\lceil \frac{D_i}{c_0}r_hf_s \rceil, L-1) \\
    h[q_i] &= h[q_i] + \frac{r^{g_i}}{D_i}
\end{align}
We set $g_i=0$ for $i=0$ (i.e., the direct-path sound source). It is easy to observe that such indexing and summation process can be done in parallel for all virtual sources to accelerate the overall simulation process. In tasks where the system is required to perform dereverberation, an early-reverberation-RIR filter is needed to serve as the target for the early reverberation component. We define the context of $[-6, 50]$ ms around the direct-path sound source as the early reverberation component:
\begin{align}
    h_e[n] = \begin{dcases*}
             h[n], & $ - \lceil\frac{6r_h f_s}{1000} \rceil \leq n - \lceil \frac{D_0}{c_0}r_h f_s \rceil \leq \lceil\frac{50r_h f_s}{1000} \rceil $ \\
             0, & otherwise
             \end{dcases*}
\end{align}
$h[n]$ and $h_e[n]$ are then passed to the same downsampling--highpass--downsampling process as mentioned in Section~\ref{sec:ISM}. 

\section{Experiment configurations}
\label{sec:config}

\subsection{Data Simulation}

The effectiveness of the proposed FRAM-RIR is evaluated on two tasks: multi-channel noisy speech
separation, and multi-channel joint noisy speech separation and dereverberation. For each utterance, we first randomly sample two speech signals from the \emph{train-clean-100} subset of Librispeech dataset \cite{panayotov2015librispeech}, and then randomly sample a noise signal from the \emph{100 Nonspeech} dataset \cite{hu2010nonspeech} and the \emph{MUSAN} dataset \cite{snyder2015musan}. The two speech utterances are then spatialized by the RIR filters, and we set the speakers to be randomly overlapped with a minimum overlap ratio of 0.5. The noise is simulated as a point-source interference spanning across the entire utterance. The signal-to-interference ratio (SIR) is randomly set within [-6, 6] dB, and the signal-to-noise ratio (SNR) is randomly set within [10, 20] dB. 

During training, the multi-channel RIR filters are simulated using different RIR simulation tools for comparison, including GPU-RIR \cite{diaz2020gpurir}, PRA-RIR \cite{scheibler2018pyroomacoustics} and the proposed FRAM-RIR. The distance between each speaker and the center of the microphone array is in the range of 0.3—6 meters. The reverberation time $T_{60}$ is randomly sampled within [0.1, 0.7] seconds. The room size is sampled with dimensions varying from 3×3×2.5 to 10×10×4 m$^3$ (length×width×height). The speakers are assumed to be static. 

For the evaluation set, a real-recorded multi-channel RIR dataset \cite{hadad2014multichannel} is adopted to generate 1,000 test utterances, where one of the recording microphone array configuration is a 8-element linear array with spacings of 4-4-4-8-4-4-4 cm. For simplicity, we use 4 of them with spacings of 4-8-4 cm. The measured $T_{60}$s are 160 ms, 360 ms and 610 ms. The sources are located on a spatial grid with azimuth range of -90$\degree$ to 90$\degree$ in 15$\degree$ steps with the distances of 1 m and 2 m to the microphone array. 

\subsection{Training Configurations}

We compare two data simulation configurations during training, namely the \emph{offline} configuration and the \emph{on-the-fly} configuration. For \emph{offline} configuration, the speech and noise signals, SNR, SIR and the corresponding RIR filters of each training sample are pre-defined and fixed for the entire training phase. In total we generate and store 100,000 RIR filters (12,500 rooms $\times$ 8 source positions). These RIR filters are convolved with speech and noise signals to generate 50,000 multi-channel mixtures. On the contrary, \emph{on-the-fly} configuration dynamically samples everything for each sample and generates unlimited numbers of utterances during training.

\subsection{Model Configurations}

To more comprehensively evaluate and compare the RIR simulation methods, we train several single-channel and multi-channel speech separation models with varying input-output configurations and model architectures. Note that we do not intend to compare the model performances but to demonstrate the effectiveness and efficiency of our proposed FRAM-RIR simulation method on a relatively wide range of common separation and denoising models. For all frequency domain models, we set the frame size and hop size to 32 ms and 8 ms, respectively. 

\begin{itemize}
    \item Single-channel bi-directional long-short term memory network (SC-BLSTM). The input feature of SC-BLSTM is the concatenation of the real and imaginary (RI) parts of the mixture spectrogram at the reference channel. There are 6 residual BLSTM layers with 256 hidden cells in each direction. The output target is the complex ratio mask (cRM) of each speaker, estimated by two fully-connected (FC) layers following the BLSTM layers.
    
    \item Single-channel band-split recurrent neural network (SC-BSRNN) \cite{luo2022bsrnn}. BSRNN is a recently proposed state-of-the-art architecture for music source separation \cite{luo2022bsrnn} and speech enhancement \cite{yu2022high}. BSRNN features with an explicit task-oriented subband splitting module and alternately performs band-level and sequence-level modeling. Here we implement a lightweight version of the original model, in which the number of BSRNN blocks is set to 8, the subband feature dimension is set to 48, and the size of hidden cells in the RNN layers is set to 96. 
    
    \item Single-channel conformer (SC-Conformer) \cite{abdulatif2022cmgan}. SC-Conformer is a powerful conformer model designed for speech enhancement in frequency domain. The input to the encoder is the combination of the real, imaginary and magnitude parts of the mixture spectrogram at the reference channel. The decoder decouples the output estimation into magnitude mask estimation and complex spectrogram refinement. While the original model contains a metric discriminator to improve the perceptual quality scores, we do not include it here to ensure a consistent training pipeline with other models. 
    
    \item Multi-channel BLSTM (MC-BLSTM). Except for the RI feature, complex interaural phase difference ($\mathbb{C}$IPD) is also concatenated as an extra input feature to provide spatial information \cite{lianwu2019multi}. 
    
    
    \item Sequential generalized Wiener filter (seq-GWF) with BLSTM and BSRNN \cite{luo2022gwf}. The sequential beamforming pipeline, i.e., a tandem pre-separation, beamforming and post-enhancement pipeline, has been recently proposed \cite{wang2021sequential, wang2020cmapping} and has shown consistent improvements compared to conventional neural beamforming pipelines. For the pre-separation module of BLSTM-seq-GWF model, the input feature and output target are set as the same with SC-BLSTM while the number of BLSTM layers is halved. Then, a GWF module is applied using the coarse estimation from the pre-separation module and the multi-channel mixture spectrograms. The RI features of the mixture spectrogram at the reference channel, pre-separated spectrograms and the spectrograms of the output of the GWF module are concatenated as the input to the post-enhancement module, which shares the same network architecture with the pre-separation module. For BSRNN-seq-GWF model, the BLSTM layers are substituted with BSRNN blocks. The total number of BSRNN blocks is 8 and the subband feature dimension is set to 32. 
    
    \item Multi-channel conformer (MC-Conformer). We simply concatenate $\mathbb{C}$IPD feature as an additional spatial feature to the single-channel input features. The output target is the same as the SC-Conformer. 
    
    \item FaSNet-TAC \cite{luo2020end}. FaSNet-TAC is an extension to the original filter-and-sum network (FaSNet) \cite{luo2019fasnet} where a transform-and-concatenate (TAC) module is applied to jointly estimate the filters for the filter-and-sum operations in all channels. We use the same configuration as the original literature, where dual-path RNN (DPRNN) blocks are used to perform sequential modeling \cite{luo2020dprnn}.
\end{itemize}
We recommend the interested readers to refer to the original literature for the details of the aforementioned models and methods.

\subsection{Training and Evaluation Configurations}
We use SNR with utterance-level permutation invariant training (uPIT) \cite{kolbaek2017multitalker} to train all the models. Adam optimizer \cite{kingma2014adam} is used with the initial learning rate of 1e-3. The learning rate is decayed by 0.98 for every 2 epochs and the early stop strategy is applied when the best model on validation data is not found in 10 consecutive epochs. All the experiments are conducted on a single GPU server with 8 NVIDIA Tesla P40 GPUs using Pytorch toolkit. For on-the-fly data simulation, the RIR filter simulation as well as the on-the-fly signal sampling and mixing is conducted in parallel using either CPU or GPU with 8 workers per data loader. 

For eavluation, we use scale-invariant SNR (SI-SNR) \cite{le2019sdr}, SNR, perceptual evaluation of speech quality (PESQ) and short-time objective intelligibility (STOI) as the metrics. To comprehensively assess the model performance, we also report performances under different azimuth ranges \cite{lianwu2019multi} (i.e., $<$15$\degree$, 15-45$\degree$, 45-90$\degree$ and $>$90$\degree$) and overlap ratios \cite{chen2020continuous} (i.e., 50-75$\%$ and $>$75$\%$).

\section{Results and analysis}
\label{sec:result}
\begin{figure*}[!t]
    \centerline{\includegraphics[width=\linewidth]{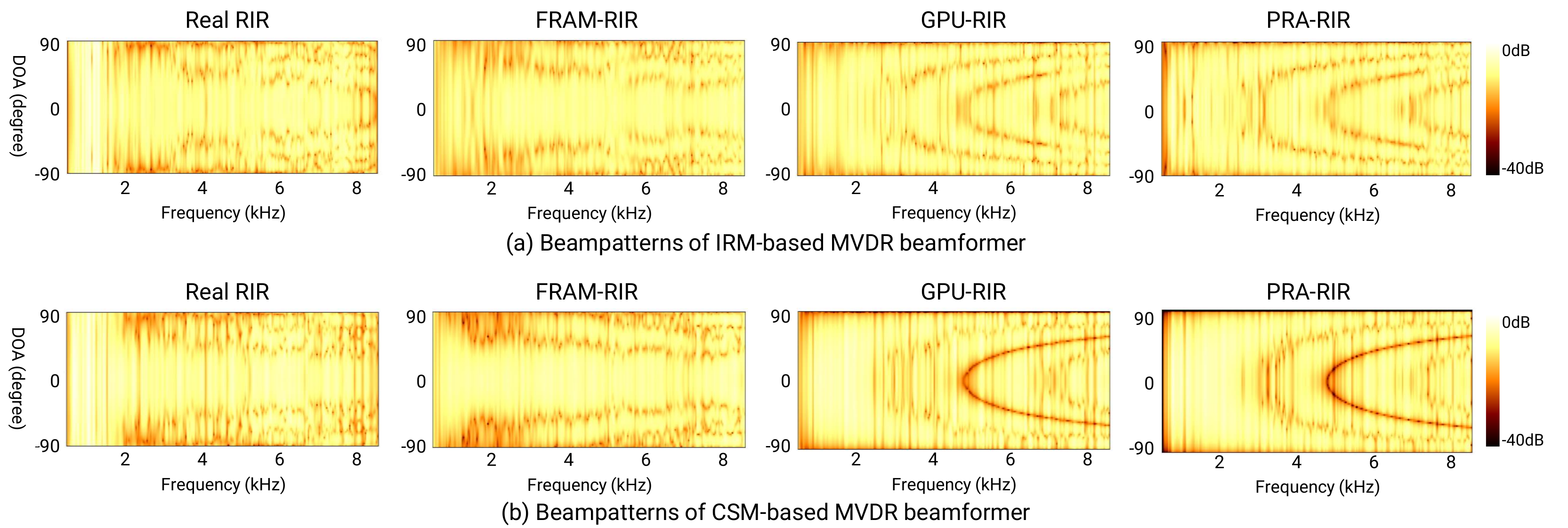}}
    \caption{Beampatterns of (a) IRM-based MVDR beamformer and (b) CSM-based MVDR beamformer with different RIR filters. Beamformer coefficients are computed upon a 2-speaker reverberant mixture spatialized with different RIR filters.  The measured $T_{60}$ of the real RIR is 360 ms, and the DOAs of the target speaker and the interfering speaker are 0$\degree$ and 90$\degree$, respectively.}
    \label{fig:beampattern}
\end{figure*}

\begin{figure*}[!t]
    \centerline{\includegraphics[width=\linewidth]{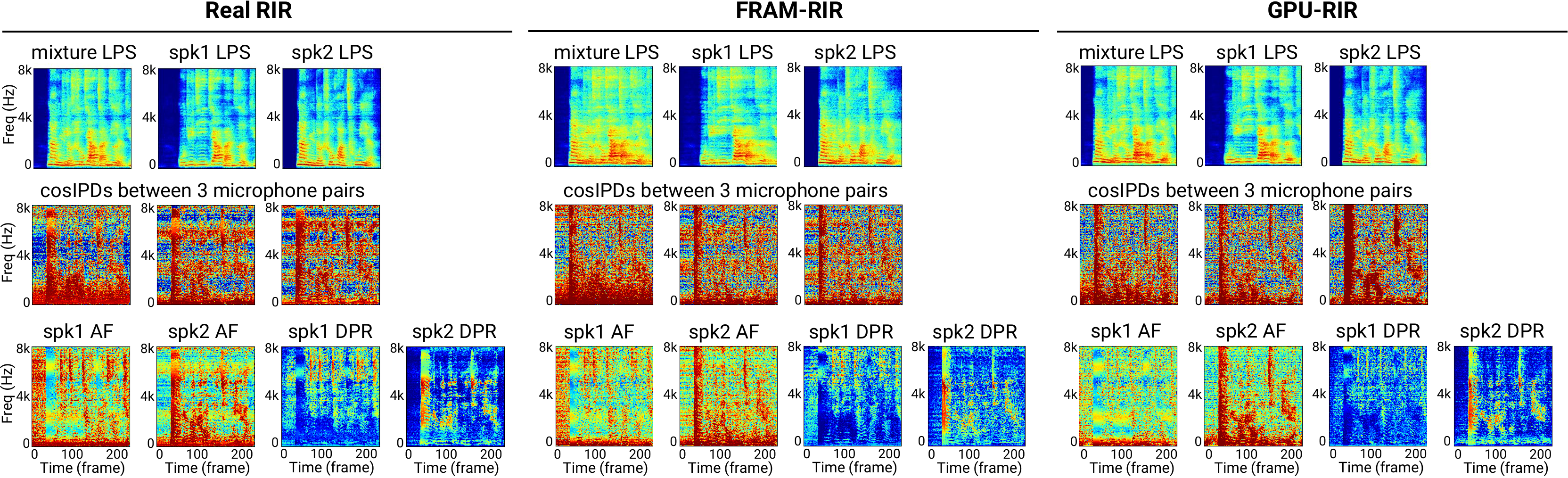}}
    \caption{Visualizations of logarithm power ratio (LPS), spatial features (cosIPDs) and directional features (AF \& DPR) extracted from 2-speaker multi-channel mixtures that are spatialized using different multi-channel RIR filters.}
    \label{fig:df}
\end{figure*}

\subsection{Verification of Directional Properties}
We first examine whether FRAM-RIR is able to preserve accurate spatial locations of the sources under the heuristic assumptions and random approximations mentioned in Section~\ref{sec:FRAM}. To do so, we visualize two spatial features: the \textit{oracle beampatterns} and the \textit{directional features}. Oracle beampatterns can be viewed as a cue for the DOAs of the direct-path sources as well as their reflections, and can then be used to verify whether the simulated RIR filters have the correct incoming direction and realistic reflection patterns. Similarly, directional features can further confirm the directional patterns of the simulated RIR filters. Since many existing models and systems take such directional features as input features, their similarity between real and simulated RIR filters can also serve as a indicator of whether the simulated RIR filters have the potential to better generalize to realistic recording conditions.

Figure~\ref{fig:beampattern} shows the oracle beampatterns generated by an ideal ratio mask (IRM)-based minimum variance distortionless response (MVDR) beamformer \cite{heymann2016neural} and an ideal complex spectral mapping (CSM) based MVDR beamformer \cite{tan2022neural,wang2020complex} with different simulated or real RIR filters. We simulate the RIR filters such that the room, microphone array and source assignments match those of the real RIR filters. We can easily observe that the beampatterns calculated from FRAM-RIR better match the real RIRs, while GPU-RIR and PRA-RIR both suffer from the sweeping echo effect of the empty shoebox-shaped room. 

Figure~\ref{fig:df} further shows the spectral and directional features, where we select the logarithm power ratio (LPS) as the spetral feature and the cosine IPD (cosIPD), angle feature (AF) \cite{chen2018efficient} and directional power ratio (DPR) \cite{gu2019neural} as the spatial features. AF and DPR are defined as follows:
\begin{align}
    \label{eq:directional}    
        \text{TPD}^{(p)}(\theta,f) &=
        \frac{2\pi f}{2(F-1)} \tau_p(\theta) \\
        \text{IPD}^{(p)}(t,f) &= \angle\mathbf{Y}^{p_1}(t,f) - \angle \mathbf{Y}^{p_2}(t,f) \\
        \text{AF}(\theta,t,f) &=\sum_{p=1}^{M} 
        \left < \text{TPD}^{(p)}(\theta,f), \text{IPD}^{(p)}(t,f) \right > \\    
        \text{DPR}(v_t,t,f) &= \frac{ \left | \textbf{w}_{v_t}^\mathsf{H}(f) \mathbf{Y}(t,f) \right |^2_F }{ \sum_{v}^{V} \left | \textbf{w}_v^\mathsf{H}(f) \mathbf{Y}(t,f) \right |^2_F }
\end{align}
where $\tau_p(\theta)$ is TDOA experienced by a sound from the target direction $\theta$ at the $p$-th microphone pair ($p_1,p_2$), $\mathbf{Y}$ denotes the multi-channel mixture spectrograms, $(\cdot)^\mathsf{H}$ denotes the complex conjugate of a matrix, $v_t$ and $v$ are the indexes of fixed beamformers that steer at the target direction and a candidate direction, respectively, and $\mathbf{w} \in \mathbb{C}^{M \times F}$ is the coefficients of the fixed beamformer. 

AF measures the similarity of the theoretical target interaural phase difference (TPDs) \cite{gu2023towards} with the observed IPDs, where each TPD is the theoretical phase difference that a unit impulse impinging from the target direction will experience at a microphone pair. When the similarity between TPDs and IPDs is larger, the probability that the corresponding T-F bin is dominated by the sound from the target direction becomes higher. DPR is defined as the power ratio between the target direction and all other directions, where the sound power from each direction can be estimated by a fixed beamformer steered at that direction, e.g., super-directive beamformer (SD-BF) \cite{gu2019neural}. It assumes spatial diversity and divides the space into fixed-resolution spatial grids. We can see from the real RIR that the directional pattern is clear and discriminative only when the DOA to compute the directional features (i.e., $\theta$ in equation 10 and $v_t$ in equation 13) matches the actual DOA of the direct sound, where the latter is formulated by the RIR patterns. By comparing the patterns of the simulated RIRs, we can find that FRAM-RIR is able to perform fine-grained DOA control and generate more realistic directional patterns.

\begin{table}[!ht]
    \centering
    \small
    \caption{Simulation speeds of 4-channel RIRs using different RIR simulation tools.}
    \begin{tabular}{c|c|c|c}
    \hline \hline
    \textbf{Method} & \textbf{\#thread} & \textbf{\#RIR (room×src)} & \textbf{speed (s)} \\
    \hline
     RIR-GEN \cite{rir_gen} & \multirow{4}{*}{1} & \multirow{4}{*}{3k (1k × 3)} & 5.665 \\  
     GPU-RIR \cite{diaz2020gpurir} & & & 0.017* \\ 
     PRA-RIR \cite{scheibler2018pyroomacoustics} & & & 0.647\\ %
     FRAM-RIR & & & 0.139 \\ 
     \hline
     GPU-RIR \cite{diaz2020gpurir} & 1 & \multirow{3}{*}{30k (10k × 3)} & 0.016* \\ 
     PRA-RIR \cite{scheibler2018pyroomacoustics} & 8 & & 0.116 \\ 
     FRAM-RIR & 8 & & 0.019 \\ 
     \hline \hline
    \end{tabular}
    \label{tab:speed}
\end{table}

\begin{table}[!ht]
    \centering
    \small
    \caption{Simulation speeds of on-the-fly batch generation using different RIR simulation tools. The batch size is set as 1.}
    \begin{tabular}{c|c|c}
    \hline \hline
    \textbf{Method} & \textbf{\#worker}   & \textbf{speed (s/batch)}\\
    \hline
     GPU-RIR \cite{diaz2020gpurir} & 1  &  0.204* \\ 
     \hline 
     PRA-RIR \cite{scheibler2018pyroomacoustics} & 1 & 2.036\\ %
     PRA-RIR \cite{scheibler2018pyroomacoustics} & 2 & 1.070\\ %
     PRA-RIR \cite{scheibler2018pyroomacoustics} & 4 & 0.594\\ %
     PRA-RIR \cite{scheibler2018pyroomacoustics} & 8 & 0.298\\ %
     \hline
     FRAM-RIR & 1 & 0.347\\ 
     FRAM-RIR & 2 & 0.165 \\ 
     FRAM-RIR & 4 & 0.087 \\ 
     FRAM-RIR & 8 & 0.054 \\     
     \hline \hline
    \end{tabular}
    \label{tab:speed2}
\end{table}

\subsection{Simulation Speed Comparison}

Besides its ability to generate more realistic RIR filters, another advantage of FRAM-RIR is that its RIR generation speed is fast enough even on CPUs so that on-the-fly data simulation becomes much easier. Here we compare and report the RIR simulation speeds of different RIR simulation tools and the mixture signal simulation speeds when the tools are used in the model training pipeline. Table~\ref{tab:speed} compares four RIR simulation tools with or without multi-threading, where GPU-RIR uses a NVIDIA Tesla P40 GPU and other methods use an Intel Xeon Platinum 8255C CPU @ 2.50 GHz. We can see that FRAM-RIR is significantly faster than other CPU-based tools when a single thread is used, and is able to achieve on par speed as GPU-RIR when multi-threading is activated\footnote{Note that GPU-RIR does not support multi-threading as by default it uses CUDA-level parallelization to accelerate the simulation.}. Table~\ref{tab:speed2} reports the on-the-fly simulation speeds of generating a batch of data with different tools using different numbers of workers in Pytorch dataloader. We sample 3 randomly source locations within the same room for each sample. It can be found that the training sample generation speed of FRAM-RIR can be faster than GPU-RIR with 2 workers and can be further accelerated by using more workers, which proves its efficiency and simplicity to use.

\subsection{Performance on Multi-channel Noisy Speech Separation and Dereverberation Tasks}

Table \ref{tab:rlt_ss} lists the results of the aforementioned single-channel and multi-channel speech separation models on the multi-channel noisy speech separation task. All models are trained on data simulated with different RIR simulation tools with either offline or on-the-fly (online) data simulation pipeline. All the evaluations are conducted on multi-channel mixtures simulated by real-recorded RIRs. The performance of IRM and IRM based oracle MVDR beamformer are reported for reference. 

We first observe that on-the-fly data simulation can always lead to a performance improvement compared to offline data simulation across all models and all RIR simulation tools, which further confirms the importance of on-the-fly training in such tasks. Moreover, we can see that GPU-RIR and FRAM-RIR perform better than PRA-RIR in most of the model architectures. Since the simulation speed for both GPU-RIR and FRAM-RIR is faster than that of PRA-RIR, one can consider replace PRA-RIR by the two counterparts for accelerated and improved on-the-fly data simulation. Comparison between GPU-RIR and FRAM-RIR shows that the performance of the models trained with the two tools varies across different model architectures and system pipelines, and overall the two tools have comparable performance. For the best-performing models, i.e., MC-Conformer and BSRNN-seq-GWF, GPU-RIR performs better especially on utterances with smaller azimuth differences, and one reason for this is that FRAM-RIR simulates more early reflection sound paths than standard ISM-based methods due to the sampling process of the propagation distances of the virtual sound sources, and it might make the directional features less discriminative when $T_{60}$ is small and the speakers are close. Such problem may be alleviated by re-balancing the distribution of the simulated data or perform room, speaker distance or $T_{60}$ dependent simulation, but we leave them as future works.

We further compare the RIR tools on a more challenging joint separation and dereverberation task, where the reverberation patterns simulated by the tools are further evaluated by examining whether the models can generalize well to real RIR filters when performing dereverberation. we use the same data and pipeline as Table~\ref{tab:rlt_ss} while change the training target from reverberant clean sources to the early reflection components of clean sources calculated by the early-reverberation-RIR filter defined in Section~\ref{sec:RIR}. Based on the results in Table~\ref{tab:rlt_ss}, we select the SC-BSRNN and BSRNN-seq-GWF models as the single-channel and multi-channel models, respectively, as they have a good trade-off between model complexity and performance. Table~\ref{tab:rlt_de} shows that the two models have similar trends in the performance on different conditions on the joint denoising, separation and dereverberation task, and FRAM-RIR still provides on par model performance as GPU-RIR.


\begin{table*}[ht!]
  \caption{Multi-channel noisy speech separation performance comparisons on test data simulated with real recorded RIRs.}
  \label{tab:rlt_ss}
  \centering
  \scalebox{0.85}{
  \begin{tabular}{l|c|c|c|c|cccc|cc|c|c|c|c}
    \hline
     \hline
    \multirow{3}{*}{\textbf{Model}} &
    \multirow{3}{*}{\textbf{\#param}} &
     &
    \multirow{3}{*}{\textbf{RIR}} &
    \multirow{3}{*}{\textbf{data}} &
    \multicolumn{7}{c|}{\textbf{SI-SDR} (dB)} &
    \multirow{3}{*}{\textbf{SNR}} &
    \multirow{3}{*}{\textbf{STOI}} &
    \multirow{3}{*}{\textbf{PESQ}} \\ 

    & & \textbf{MAC}  & & &\multicolumn{4}{c|}{Azm Difference} & \multicolumn{2}{c|}{Overlap Ratio} & \multirow{2}{*}{\textbf{Avg}} && \\
    & & (G/s) & & & $<$15\degree & 15-45\degree & 45-90\degree & $>$90\degree & 50-75\% & $>$75\% &  & (dB) & \\
    \hline\hline
    
    Mixture & - & - & -  & - & -0.4 & -0.4 & -0.4 & -0.4 & -0.4 & -0.4 & -0.4 & -0.4 & 0.52 & 1.10 \\    
    IRM & - & - & - & - & 11.1 & 11.0 & 11.2 & 11.1 & 11.3 & 10.9 & 11.1 & 11.4 & 0.89 & 1.76 \\
    IRM-MVDR & - & - & - & - & 4.0 & 4.5 & 6.2 & 6.6 & 5.8 & 5.8 & 5.8 & 5.8 & 0.70 & 1.28 \\
    \hline \hline
    \multirow{6}{*}{SC-BLSTM} & \multirow{6}{*}{5.6M} & \multirow{6}{*}{0.74} & GPU & \multirow{3}{*}{offline} &  4.8 & 3.9 & 4.2 & 4.3 & 4.3 & 4.1 & 4.2 & 5.8 & 0.67 & 1.24 \\  
    & & & PRA &  & 3.8 & 3.0 & 3.3 & 3.4 & 3.5 & 3.1 & 3.3 & 5.1 & 0.66 & 1.24 \\   
    & & &FRAM &  &  4.6 & 3.6 & 4.1 & 4.2 & 4.5 & 3.6 & 4.1 & 5.7 & 0.68 & 1.27 \\    
    \cline{4-15} 
    & && GPU  & \multirow{3}{*}{online} &  6.3 & 5.5 & 5.8 & 6.1 & 6.1 & 5.6 & 5.8 & 7.0 & 0.72 & 1.30 \\   
    & &&PRA  & &   5.2 & 4.5 & 4.9 & 5.1 & 5.1 & 4.6 & 4.9 & 6.3 & 0.68 & 1.27  \\   
    & && FRAM & &  6.5 & 5.5 & 5.8 & 6.1 & 6.2 & 5.5 & 5.9 & 7.0 & 0.71 & 1.31 \\
    \hline
    \multirow{3}{*}{MC-BLSTM} & \multirow{3}{*}{5.8M} & \multirow{3}{*}{0.75} & GPU & \multirow{3}{*}{online} & 4.4 & 6.7 & 8.9 & 9.7 & 8.8 & 8.0 & 8.4 & 9.1 & 0.82 & 1.46 \\ 
    & && PRA &  & 4.9 & 7.1 & 9.5 & 10.2 & 9.3 & 8.5 & 8.9 & 9.5 & 0.84 & 1.49\\     
    &&& FRAM & & 5.3 & 7.3 & 9.7 & 10.3 & 9.5 & 8.6 & 9.1 & 9.7 & 0.83 & 1.51 \\      
    \hline \hline
    \multirow{6}{*}{BLSTM-seq-GWF} & \multirow{6}{*}{7.0M} & \multirow{6}{*}{9.71} & GPU & \multirow{3}{*}{offline} & 5.4 & 4.7 & 5.1 & 7.0 & 6.1 & 5.9 & 6.1 & 7.2 & 0.74 & 1.34 \\  
    & && PRA  & & 6.0 & 5.1 & 6.7 & 7.6 & 6.8 & 6.3 & 6.5 & 7.7 & 0.75 & 1.37 \\   
    & && FRAM  & & 6.3 & 5.7 & 7.2 & 7.9 & 7.2 & 6.8 & 7.0 & 8.0 & 0.76 & 1.38  \\    
    \cline{4-15} 
    &&&  GPU & \multirow{3}{*}{online} & 10.1 & 9.8 & 10.7 & 11.0 & 11.0 & 10.0 & 10.5 & 11.0 & 0.84 & 1.62 \\
    &&&  PRA  &  & 8.4 & 7.7 & 9.3 & 9.7 & 9.2 & 8.7 & 9.0 & 9.6 & 0.81 & 1.48  \\   
    &&&  FRAM &  & 10.3 & 9.6 & 10.7 & 11.0 & 10.9 & 10.1 & 10.5 & 11.0 & 0.85 & 1.62 \\
    \hline \hline
    \multirow{6}{*}{FaSNet-TAC} & \multirow{6}{*}{2.8M} & \multirow{6}{*}{10.05} & GPU & \multirow{3}{*}{offline} & 4.4 & 4.6 & 7.2 & 9.0 & 7.2 & 6.9 & 7.0 & 8.0 & 0.76 & 1.37 \\   
    &&& PRA  & & 2.6 & 3.0 & 6.3 & 8.1 & 6.0 & 5.8 & 5.9 & 7.1 & 0.72 & 1.32  \\   
    &&& FRAM  & &   6.3 & 4.9 & 5.4 & 5.8 & 5.8 & 5.1 & 5.5 & 6.8 & 0.71 & 1.31 \\ 
    \cline{4-15} 
    & & & GPU & \multirow{3}{*}{online} & 4.9 & 5.4 & 7.9 & 9.3 & 7.9 & 7.2 & 7.6 & 8.4 & 0.77 & 1.37 \\   
    & & &PRA  &  &  4.2 & 4.6 & 6.8 & 7.6 & 6.6 & 6.2 & 6.4 & 7.4 & 0.73 & 1.32   \\   
    & & &FRAM  &  &  6.9 & 6.3 & 7.6 & 8.6 & 7.8 & 7.2 & 7.6 & 8.4 & 0.77 & 1.39 \\     
    \hline \hline

    \multirow{3}{*}{SC-Conformer} & \multirow{3}{*}{1.8M} & \multirow{3}{*}{31.62} & GPU & \multirow{3}{*}{online} & 13.2 & 12.4 & 13.2 & 13.0 & 13.3 & 12.4 & 12.9 & 13.2 & 0.92 & 1.91 \\
    &&& PRA &  & 12.7 & 12.3 & 12.8 & 12.7 & 13.0 & 12.2 & 12.6 & 12.9 & 0.91 & 1.87 \\
    &&& FRAM & & 13.2 & 12.4 & 13.1 & 13.0 & 13.3 & 12.4 & 12.9 & 13.1 & 0.90 & 1.89 \\
    \hline

    \multirow{3}{*}{MC-Conformer}  & \multirow{3}{*}{1.8M} & \multirow{3}{*}{31.63} & GPU & \multirow{3}{*}{online} & 13.3 & 13.3 & 14.2 & 14.5 & 14.4 & 13.6 & 14.0 & 14.2 & 0.86 & 2.05 \\
    &&& PRA & & 12.8 & 12.9 & 14.0 & 14.3 & 14.1 & 13.3 & 13.7 & 13.9 & 0.94 & 2.06\\
    &&& FRAM & & 11.5 & 12.6 & 13.8 & 13.9 & 13.8 & 12.9 & 13.4 & 13.6 & 0.93 & 1.99  \\
    \hline \hline
    \multirow{3}{*}{SC-BSRNN} & \multirow{3}{*}{3.3M} & \multirow{3}{*}{6.38} & GPU & \multirow{3}{*}{online} &  12.4 & 11.8 & 12.4 & 12.4 & 12.6 & 11.8 & 12.2 & 12.5 & 0.91 & 1.95 \\     
     &&& PRA & & 12.0 & 11.3 & 11.9 & 11.9 & 12.2 & 11.3 & 11.7 & 12.1 & 0.90 & 1.94 \\ 
     &&& FRAM & & 12.5 & 11.8 & 12.4 & 12.3 & 12.6 & 11.8 & 12.2 & 12.5 & 0.91 & 1.96 \\ 
   \hline
   \multirow{3}{*}{BSRNN-seq-GWF} & \multirow{3}{*}{3.7M} & \multirow{3}{*}{14.62} & GPU & \multirow{3}{*}{online} & 13.6 & 13.4 & 14.2 & 14.3 & 14.3 & 13.6 & 14.0 & 14.2 & 0.93 & 2.13 \\
   &&& PRA & & 12.8 & 12.6 & 13.7 & 13.9 & 13.8 & 13.0 & 13.4 & 13.6 & 0.92 & 2.04 \\
   &&& FRAM & & 13.2 & 12.9 & 13.9 & 14.2 & 14.1 & 13.3 & 13.7 & 13.9 & 0.92 & 2.07 \\
   \hline \hline

 \end{tabular}}
\end{table*}

\begin{table*}[ht!]
  \caption{Multi-channel joint noisy speech separation and dereverberation performance comparisons on test data simulated with real recorded RIRs.}
  \label{tab:rlt_de}
  \centering
  \scalebox{0.9}{
  \begin{tabular}{l|c|c|c|cccc|cc|c|c|c|c}
    \hline
     \hline
    \multirow{3}{*}{\textbf{Model}} &
    \multirow{3}{*}{\textbf{\#param}} &
    \multirow{3}{*}{\textbf{RIR}} &
    \multirow{3}{*}{\textbf{data}} &
    \multicolumn{7}{c|}{\textbf{SI-SDR (dB)}} &
    \multirow{3}{*}{\textbf{SNR}} &
    \multirow{3}{*}{\textbf{STOI}} &
    \multirow{3}{*}{\textbf{PESQ}} \\ 
    
    & & & &\multicolumn{4}{c|}{Azm Difference} & \multicolumn{2}{c|}{Overlap Ratio} & \multirow{2}{*}{\textbf{Avg}} && \\
    & & & & $<$15\degree & 15-45\degree & 45-90\degree & $>$90\degree & 50-75\% & $>$75\% &  && \\
    \hline\hline
    
    Mixture & -  & -  & -  & -1.8 & -1.6 & -1.7 & -1.5 & -1.6 & -1.6 & -1.6 & -0.9 & 0.50 & 1.08 \\    
    IRM & - & -  & -  & 7.0 & 7.3 & 7.3 & 7.6 & 7.6 & 7.2 & 7.4 & 8.3 & 0.85 & 1.53 \\
    IRM-MVDR & -  & - & -  & 2.6 & 3.0 & 4.3 & 4.9 & 4.1 & 4.1 & 4.1 & 5.1 & 0.67 & 1.23 \\
    \hline\hline

    \multirow{3}{*}{SC-BSRNN}  & \multirow{3}{*}{3.3M} & GPU & \multirow{3}{*}{online} & 7.5 & 7.4 & 7.6 & 7.9 & 7.9 & 7.4 & 7.7 & 8.7 & 0.84 & 1.58 \\
    & & PRA & & 7.5 & 7.0 & 7.4 & 7.8 & 7.8 & 7.1 & 7.5 & 8.5 & 0.84 & 1.58 \\        
    & & FRAM & & 7.7 & 7.5 & 7.8 & 8.1 & 8.2 & 7.5 & 7.8 & 8.9 & 0.85 & 1.60 \\
    \hline\hline
    \multirow{3}{*}{BSRNN-seq-GWF}  & \multirow{3}{*}{3.7M} & GPU & \multirow{3}{*}{online} & 8.9 & 9.1 & 9.5 & 10.2 & 9.9 & 9.3 & 9.6 & 10.4 & 0.88 & 1.82 \\
    & & PRA & & 8.5 & 8.7 & 9.2 & 9.9 & 9.5 & 9.0 & 9.3 & 10.1 & 0.88 & 1.80 \\        
    & & FRAM & & 8.8 & 8.9 & 9.5 & 10.1 & 9.8 & 9.2 & 9.5 & 10.4 & 0.88 & 1.78  \\
    \hline \hline
 \end{tabular}}
\end{table*}


\subsection{$T_{60}$ Curriculum Training}

\begin{figure}[!ht]
    \centerline{\includegraphics[width=0.9\linewidth]{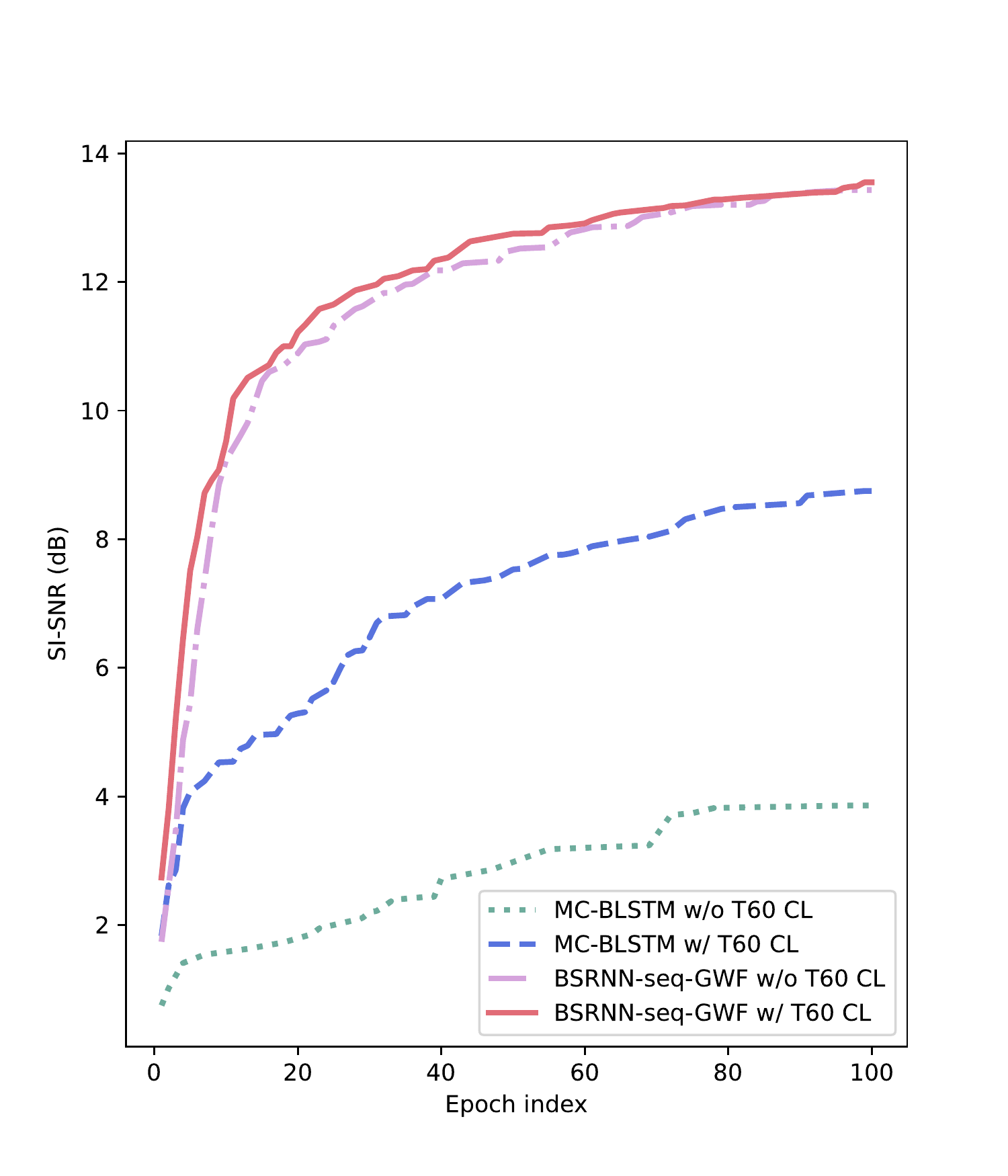}}
    \caption{SI-SNR performance (dB) on test data versus training epoch when $T_{60}$ curriculum learning is applied (w/ $T_{60}$ CL) or not (w/o $T_{60}$ CL) to MC-BLSTM and BSRNN-seq-GWF models.}
    \label{fig:t60_cl}
\end{figure}

Certain applications may require the model to be robust in scenarios with a wide range of $T_{60}$, e.g., from small bedrooms or meeting rooms to large classrooms or concert halls. We empirically find that directly training models with such a large $T_{60}$ range may lead to suboptimal performance and slower convergence speed, and one possible reason might be that large $T_{60}$ may lead to more complicated patterns for directional features and makes the model hard to properly utilize the spectral and spatial features at the early stages of training. Here we validate \textit{$T_{60}$ curriculum training}, a training technique proposed to mitigate the issue \cite{aralikatti2021improving}, on the on-the-fly training pipeline with FRAM-RIR. Given a maximum $T_{60}$ value that the on-the-fly data simulation pipeline is required to cover, we start the model training by sampling $T_{60}$ within a relatively smaller range, and gradually increase the upper bound of the range as the training continues. We empirically find that by setting the initial $T_{60}$ range to [50, 100] ms and increase the upper bound by 50 ms for every subsequent epoch until it reaches the pre-defined maximum $T_{60}$ can effectively accelerate the model convergence and improve the performance. Figure~\ref{fig:t60_cl} shows the performance of two systems, the MC-BLSTM model and the BSRNN-seq-GWF model, trained with or without $T_{60}$ curriculum training. We train both models for a maximum of 100 epochs and report the SI-SNR score on the test set at different epochs. It can be observed that the convergence speed for both models can be effectively accelerated especially at early epochs, and the performance improvement for MC-BLSTM is significant when $T_{60}$ curriculum training is applied. The results indicate that $T_{60}$ curriculum training can be an effective supplement within the on-the-fly training pipeline.

\section{Conclusion and future works}
\label{sec:conclusion}
In this paper, we proposed fast random approximation of multi-channel room impulse response (FRAM-RIR), a method to promptly simulate realistic RIR filters for data augmentation purpose in the training of speech processing systems. FRAM-RIR can not only simulate RIR filters that better mimic the patterns of real RIR filters than existing tools, but is also significantly faster than other CPU-based methods with a single thread and faster than GPU-accelerated methods with multi-threading. Experiment results showed that FRAM-RIR enabled fast on-the-fly training with on par or better performance than other RIR simulation tools across a wide range of models in the multi-channel noisy speech separation and dereverberation tasks. Moreover, a $T_{60}$ curriculum training strategy was proposed to accelerate the convergence speed during training phase and to improve the model performance. Future works include the application and verification of FRAM-RIR in other types of speech processing tasks and the extension of its design and application towards more complicated scenarios, e.g., enabling RIR simulation with moving sources and supporting microphone modeling.

\bibliographystyle{IEEEbib}
\bibliography{main}

\end{document}